\documentstyle[preprint,aps,prb]{revtex}

\begin{document}
\title{Surveying the solar system by measuring angles and times:
from the solar density to the gravitational constant}

\author{Klaus Capelle}
\date{\today}
\address{Instituto de F\'{\i}sica de S\~ao Carlos,
Universidade de S\~ao Paulo, Caixa Postal 369, S\~ao Carlos, 13560-970 SP,
Brazil}
\maketitle
\begin{abstract}
A surprisingly large amount of information on our solar system can be gained 
from simple measurements of the apparent angular diameters of the sun and the 
moon. This information includes the average density of the sun, the distance 
between earth and moon, the radius of the moon, and the gravitational constant.
In this note it is described how these and other quantities can be obtained by 
simple earthbound measurements of angles and times only, without using any 
explicit information on distances between celestial bodies. The pedagogical 
and historical aspects of these results are also discussed briefly.
\end{abstract}


\newcommand{\be}{\begin{equation}}
\newcommand{\ee}{\end{equation}}
\newcommand{\bi}{\bibitem}

\newpage 
\section{Introduction}
\label{intro}

All standard textbooks of astronomy and classical mechanics show how the 
great insights of Kepler and Newton can be used to determine the geometry 
of our solar system and the physical properties of its constituent bodies
(Refs.~\onlinecite{carroll,knudsen,lang} are representative examples).
Such calculations are normally based on simple relations between masses 
and distances, which follow directly from Newton's or Kepler's laws.
The purpose of the present note is to discuss a little noticed similar 
relation that allows us to determine the densities and other properties of 
celestial bodies, as well as the gravitational constant, from entirely 
earthbound and very simple observations of angles and times.

Section \ref{sun} of this paper shows how the average solar density can be 
obtained from knowing nothing more than the apparent angle under which the 
sun appears as seen from the earth and the duration of a year. 
Section \ref{moon} applies the same idea to the moon. Although the
details are slightly more complicated in this case, it still proves
possible to calculate the moon's average distance from the earth, and its 
radius, by measuring the density of the earth.
Section \ref{gamma} shows how the relations derived can also be 
used to calculate the gravitational constant (which was not known to
Newton) by using only data available at Newton's time. 

Although, of course, nothing new is added to current research in astronomy
by these considerations, is seems that the resulting curious relations 
are well suited to stimulate a student's interest in the subject.
They might also be interesting from the point of view of the history of
science, because they show how Newton could, e.g., have obtained 
the gravitational constant long before the celebrated Cavendish 
experiment.\cite{cavendish}
Another aspect of the relations discussed here is that they serve as vivid
illustrations of how indirect measurements, together with the assumed
universal validity of the laws of physics, allow us to obtain information
on quantities that are completely inaccessible for direct measurements.
These pedagogical and historical issues are taken up again in the final
Section \ref{discussion} of this little note.

\section{Physical properties of the sun}
\label{sun}
In order to derive an expression for the average density of the sun,
$\rho_s$, we
start by writing this density as the ratio of total mass to total volume,
\be
\rho_s = \frac{M}{V} = \frac{3M}{4\pi R_s^3},
\label{rhodef}
\ee
where the sun is assumed to be a perfect sphere of radius $R_s$.
The mass can be eliminated from this equation by equating the centripetal
force the earth experiences on its (approximately circular) orbit with the 
gravitational attraction of the sun, which is assumed to be so much
heavier than the earth that the center of mass of the system earth-sun
coincides with that of the sun alone. Hence,
\be
-m\omega^2 r = - G \frac{mM}{r^2},
\label{forces}
\ee
where $m$ is the earth's mass, $G$ the gravitational constant, and
$r$ the distance from earth to sun. Expressing the angular velocity 
of the earth, $\omega$, in terms of its rotation period, $T=2\pi/\omega$, 
and solving Eq.~(\ref{forces}) for $M$, one readily finds
\be
M=\frac{4\pi^2 r^3}{G T^2},
\label{meq}
\ee 
which is essentially Kepler's third law. 

As seen from Fig.~\ref{fig1} the distance $r$ between earth and sun is 
related to the angle under which the sun appears from the center of the 
earth by
\be
\sin(\beta)=\frac{R_s}{r},
\label{betadef}
\ee
where $2 \beta$ is the apparent angular diameter of the sun. Since the radius 
of the earth is much smaller than the distance $r$ between earth and sun, this
equation still holds approximately when $\beta$ is measured from the surface 
of the earth, and not its center. (In technical terms this means that 
surface parallax can be neglected.) Since the solar radius $R_s$ is also much
smaller than $r$, $\beta$ is a very small angle and we could replace 
$\sin(\beta)$ by $\beta$ with negligible error, but for generality and future 
applicability to other systems we keep the trigonometric functions here and 
below.

Substituting Eq.~(\ref{betadef}) into Eq.~(\ref{meq}) and the result into 
Eq.~(\ref{rhodef}) one finds that the solar radius $R_s$ cancels, and what 
remains is an expression for $\rho_s$ that does not explicitly involve any 
distances,
\be
\rho_s=\frac{3\pi}{G T^2 \sin^3(\beta)}.
\label{density}
\ee
To the best of the author's knowledge Eq.~(\ref{density}) does not
appear explicitly in any of the standard English-language textbooks 
(although, as the author learned after completing this work, its derivation 
is asigned as a problem in Ref.~\onlinecite{nussenzveig}). 
The remainder of this note is dedicated to exploration of a few interesting 
consequences of this result.

A remarkable feature of Eq.~(\ref{density}) is that all quantities refering
explicitly to diameters and distances have disappeared. The remaining 
quantities, $T$ and $\beta$, are accessible via purely earthbound and very 
simple measurements.
$T$ is simply the duration of a year, and $2\beta$, the angle subtended
by the sun as seen from the earth, is found to be about half a degree.
This value varies slightly during a year because the earth's orbit is not
exactly circular, but this variation is neglected below, where 
$\beta=0.25^o$ is adopted for convenience. 
(Suggestions how to measure $\beta$ using simple equipment are 
made in Sec.~\ref{discussion}, below). 
The gravitational constant $G=6.7 \times 10^{-11}\, {\rm m^3 s^{-2} kg^{-1}}$
must be known to evaluate Eq.~(\ref{density}). This does not impede
its use in the classroom, but would have had interesting consequences if 
Newton had known that equation. Some of these consequences are 
explored in the following sections.
Plugging the above numbers in Eq.~(\ref{density}) one obtains 
\be
\rho_s = 1.7 \times 10^3\, {\rm kg\, m^{-3}} = 1.7\, {\rm g\, cm^{-3}},
\label{dnumber}
\ee
where the second equation expresses the result in units more common in
astronomy. 

The curious little formula (\ref{density}) never fails to surprise
students and even more mature scientists.
The surprise is not really that the value found in Eq.~(\ref{dnumber})
agrees rather well with the literature value\cite{handbook} 
$\rho_s =1.4\, {\rm g\, cm^{-3}}$ (the main source of error doubtlessly being 
the imprecise measurement of the angular diameter of the sun\cite{footnote1}), 
but that it has been obtained without measuring any distances.
It thus provides us with a rather different type of information about our 
solar system than do the classical measurements of distances and diameters,
which usually treat celestial objects as point masses without internal 
properties. While these determine the {\em scale} of the solar system, 
knowledge of the density allows us to draw conclusions concerning the 
internal {\em composition} of its constituents.

If one, in an admittedly arbitrary way, 
takes solid iron as representative of the earth's metals
($\rho_{\rm Fe}=7.87\, {\rm g\, cm^{-3}}$),\cite{handbook}
silicon dioxide as representative for rocks and sand 
($\rho_{\rm SiO_2}=2.65\, {\rm g\, cm^{-3}}$),\cite{handbook}
and both of these as representative for the composition of the earth as
a whole, it immediately follows that the sun is {\em not} composed 
primarily of these solid materials, and thus of a different physical nature 
than the earth. Furthermore, by comparing with the densities of other solids, 
liquids, and gases one concludes that the sun is, due to its low average 
density, most likely not solid at all.\cite{footnote2} 

Newton himself, by using a related but much more complicated method, 
based on observations of the orbit of the planets around the sun
and of the moon around the earth,
arrived at a very similar conclusion. In the {\em Principia} he writes:
{\em The sun, therefore, is a little denser than Jupiter, and Jupiter than 
Saturn, and the earth four times denser than the sun; for the sun, by its 
great heat, is kept in a sort of rarefied state}.\cite{newton1}
A little later he provides his estimate of the density of the earth:
{\em Since, therefore, the common matter of our earth on the surface thereof
is about twice as heavy as water, and a little lower, in mines, is found
about three, or four, or even five times heavier, it is probable that the
quantity of the whole matter of the earth may be five or six times greater 
than if it consisted all of water; ...}.\cite{newton2}

This estimate of the density of the earth, five or six times that of water,
is remarkably close to the modern value\cite{handbook} 
of $5.5\, {\rm g\, cm^{-3}}$.
Together with the factor of four by which, according to him, the earth
is denser than the sun, one finds that the solar density is inbetween about
$1.3 \, {\rm g\, cm^{-3}}$ and $1.5 \,{\rm g\, cm^{-3}}$, which is even closer 
to the modern value than that found from Eq.~(\ref{density})
(but obtained with considerably more labour). 

\section{Distance and radius of the moon}
\label{moon}

It is tempting to apply Eq.~(\ref{density}) to the moon as well. 
After all, the moon's apparent angular diameter is almost exactly the same
as that of the sun (as evidenced by solar eclipses or direct measurement), 
and its orbital period is also well known.
One could thus determine the average density of ... of what? Of the earth
or the moon? The answer is that this tentative procedure does not 
directly determine the average density of either of these bodies. 
We can again use Fig.~\ref{fig1}, which we already used in our determination
of the solar density, to understand why. Quite generally, this figure depicts 
angles and distances characterizing the geometry of the revolution of a 
lighter body orbiting around a heavier one. However, when we are dealing with
the system earth-moon instead of sun-earth, our point of view has shifted from 
the orbiting to the central body. This means that Eq.~(\ref{betadef}) is not 
directly applicable here: the mass $M$ in Eq.~(\ref{rhodef}) is that of the 
central body, so that the angular diameter and the density in 
Eq.~(\ref{betadef}) are that of the central body as seen from the accompanying
body. 

This is not the end of the story, though. Eq.~(\ref{density}) determines 
the average density of the central body, i.e., in the case of the system 
earth-moon that of the earth. This density is reasonably well approximated 
by that of the typical materials mentioned above, and can be obtained 
without requiring any input from celestial mechanics. This enables us to 
turn the argument around and obtain information on the moon from 
Eq.~(\ref{density}), by treating the earth's density as a known quantity. 
Let us denote the apparent angle of the moon as seen from the earth
by $2\alpha_m$ and that of the earth as seen from the moon by $2\beta_e$.
This latter angle seems hard to obtain in preastronautical times, but
from Fig. \ref{fig1}, reinterpreted now as depicting the system
earth-moon, it follows that 
\be
\sin(\beta_e) = \frac{R_e}{R_m} \sin(\alpha_m),
\label{angles}
\ee
where $R_e$ and $R_m$ denote the radii of earth and moon, respectively. 
This relation allows us to eliminate the unknown angle $\beta_e$ from the 
equations. Eq.~(\ref{density}), rewritten for the system earth-moon, reads
\be
\rho_e=\frac{3\pi}{G T_m^2 \sin^3(\beta_e)},
\ee
where $T_m$ is the duration of the orbital period of the moon (about $27$
days), and $\rho_e$ the average density of the earth.
From this expression we obtain with the help of Eq.~(\ref{angles}) that
\be
\rho_e = 
\frac{3\pi}{G T_m^2}\frac{R_m^3}{R_e^3} \frac{1}{\sin^3(\alpha_m)}.
\ee
Assuming again that the average density of the earth is close to the average 
of that of the typical materials iron and silicon dioxide, given above, 
and treating the radius of the earth as a known quantity (which it certainly
was at Newton's times), we can calculate the radius of the moon from this 
equation.
Putting in the numbers yields
\be
R_m=\left(\frac{G T_m^2 \rho_e}{3\pi}\right)^{1/3}R_e \sin(\alpha_m)
\approx 1600 \,{\rm km},
\ee
which is to be compared with the literature value\cite{lang} of 
$1738\,{\rm km}$.
The difference between both values has several sources.
First, there is the rather arbitrary choice of using the average density of 
iron and silicon dioxide to represent that of the earth.
Second, the angular diameter of the moon is only imprecisely known, and
also changes slightly during the course of a month.
Third, the parallax resulting from the fact that the observer is on the
surface of the earth and not at its center, has been neglected.
Finally, the derivation of Eq.~(\ref{density}) is only correct for circular 
orbits around a stationary central body, an assumption which is less well 
satisfied for the motion of the moon around the earth than for that of the 
earth around the sun. It can be instructive to discuss with students which 
of these is the dominant source of error. Discussions of how to improve on 
some of these aspects of the above procedure can make for rewarding science 
projects of high-school level students, or be used as homework problems
for more advanced ones.

Since the {\it ratios} between many distances in the solar system were already 
known to Newton and his contemporaries, the determination of a single
{\it absolute value}, like $ R_m$, enables one to calculate the other
distances explicitly. As an example, we can now work backwards from the 
counterpart to Eq.~(\ref{betadef}) for the system earth-moon, 
\be
\sin(\alpha_m)=R_m/r_{e-m},
\ee
where $r_{e-m}$ stands for the distance earth-moon, and find 
$ r_{e-m}\approx 3.7 \times 10^5 \, {\rm km}$. The literature 
value\cite{handbook} for the average distance is 
$3.8 \times 10^5 \,{\rm km}$. Students may find it rewarding to reflect
about how it was possible to come this close to today's value for the 
distance of the moon by measuring the density of the earth, and what margin
of error the various approximations made imply for the final value.

\section{The gravitational constant}
\label{gamma}

Newton himself did not know the numerical value of the gravitational
constant. Following the style of reasoning common in his days he expressed 
his results in terms of {\it ratios} between masses, distances and
densities. From such ratios the constant prefactor $G$ of course
disappears. Hence, in Newton's days there seems to have been little
interest in determining this and similar universal prefactors.
The first reasonably precise determination of $G$ was made possible 
in 1798 by Henry Cavendish, more than hundred years after Newton developed
his theory of gravitation, and even Cavendish's experiment was not 
explicitly recognized as a determination of $G$ until another hundred 
years later.\cite{cavendish}

Newton's lack of knowledge concerning the value of $G$ is particularly 
surprising in view of the fact that he could, for example, have obtained 
this value from the equation of motion for a test particle of mass $m_t$ in 
the gravitational field of the earth,
\be
m_tg = G \frac{m_tM_e}{R_e^2},
\ee
where $g$ is the gravitational acceleration at the earth's surface.
By expressing the mass of the earth, $M_e$, in terms of its radius and density,
and using $g=9.81\, m s^{-2}$ one readily obtains $G$. 

Apparently Newton did not do this simple calculation.
He did, however, work hard to obtain his estimate of the density of the sun
quoted above.
This estimate, in conjunction with Eq.~(\ref{density}), opens up another
path for determining the gravitational constant, which would have also
been available to Newton.
By solving Eq.~(\ref{density}) for $G$, substituting the numerical
values for $T$ and $\beta$ (which were both available at Newton's times), 
and using Newton's own estimate for $\rho_s$, quoted in Sec.~\ref{sun}, 
one finds
\be
G = \frac{3\pi}{\rho_s T^2 \sin^3(\beta)}
\approx 8.1\times 10^{-11}\, {\rm m^3 s^{-2} kg^{-1}}, 
\ee
which is surprisingly close to the modern value of 
$6.7 \times 10^{-11}\, {\rm m^3 s^{-2} kg^{-1}}$, and could have been obtained
by Newton and his contemporaries, or later by Cavendish, without having to 
perform difficult measurements of the mutual attraction of masses. 
\cite{cavendish}

\section{discussion}
\label{discussion}

From a pedagogical point of view, the above calculations demonstrate,
in a very simple case, the power of physics. 
Measuring nothing more than the duration of a year 
and the apparent angular diameter of the sun we can obtain the solar 
density, a number which is not related to these two quantities in any obvious
way.
Surprises like this may be a valuable pedagogical tool, since they illustrate
vividly how the universal validity of the laws of physics allows us to obtain 
information on properties of nature which are totally inaccessible by
means of direct measurement.
Apart from this more philosophical aspect, the above little calculations are
also well suited as classroom exercises in an introductory astronomy 
course on the undergraduate or high-school level, since they require nothing 
but the most basic classical mechanics, a measurement of the angular diameter 
of the sun, and simple algebra. 

This angle can be measured directly if a telescope with
a cross wire eye piece and a solar filter is available.
Using the fact that the sun traverses $360$ degrees in one day,
and measuring the time it takes the sun to traverse its own apparent diameter, 
one immediately obtains $2 \beta $. 
Due to the great intensity of the sunlight such a direct measurement is 
somewhat dangerous and it may be preferable to image the sun with a lense of 
known focal length instead. 
Alternative ways of measurement, accessible to high-school or undergraduate
students, include estimating the angular diameter of the 
sun from the duration of a sunset, or directly measuring the angular 
diameter of the moon. As pointed out above, this diameter is almost identical 
to that of the sun, a fact that is most impressively demonstrated by showing
pictures of solar eclipses. 
A discussion with the students of the advantages and disadvantages
of the various procedures for obtaining $2 \beta $ can be very instructive.

From the point of view of history of science, the above considerations 
demonstrate that Newton (or his successors) could have obtained more detailed 
information concerning the nature of our solar system than they seem to have 
done. These remarks are in no way meant to diminish Newton's
tremendous intellectual achievements, but they show that purely earthbound 
and very simple mesurements allow to obtain much more detailed information
than is often thought. What makes this possible is precisely the
generality of Newtonian mechanics, the universal validity of which implies
relations between quantities measured on earth and quantities pertaining
to other celestial bodies.
\\ \\
\noindent
{\bf Acknowledgments}\\
It is a pleasure to thank L.~N.~Oliveira, V.~L.~Libero, and S.~Ragusa 
for interesting discussions on the subject matter of this note.

\newpage

\begin{figure}
\caption{Geometry of the motion of an orbiting lighter body of radius
$R_e$, revolving about a heavier stationary body of radius $R_s$, at 
distance $r$. In Sec.~\ref{sun} the heavier body is the sun and the lighter 
the earth, while in Sec.~\ref{moon} they are earth and moon, respectively. 
The angle $\beta$ is half the apparent angular diameter of the central body, 
as seen from the orbiting body (neglecting parallax), while $\alpha$ is that 
of the orbiting body, as seen from the central body.} 
\label{fig1}
\end{figure}

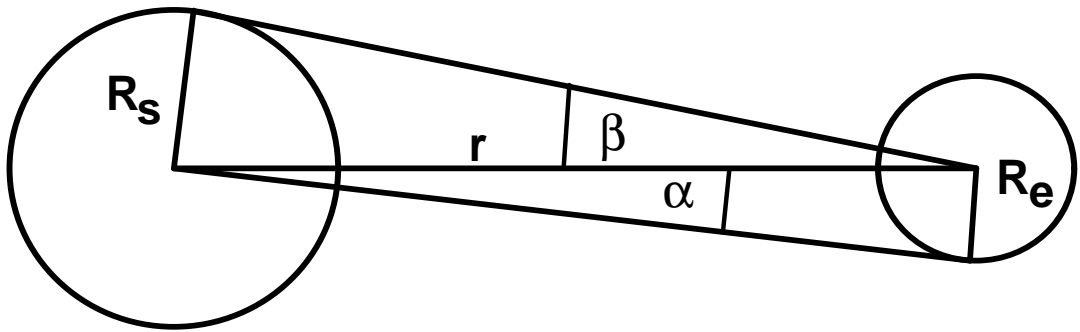
\end{document}